\documentclass[pra, showpacs,onecolumn,eqsecnum,superscriptaddress]{revtex4}
\usepackage{makeidx}
\usepackage{eurosym}
\usepackage{amssymb}
\usepackage{amsfonts}
\usepackage{amsmath}
\usepackage{setspace}
\usepackage[english]{babel}
\usepackage{graphicx}
\usepackage[applemac]{inputenc}
\usepackage{color}
\usepackage{ifthen}
\usepackage{float}
\usepackage{verbatim}
\usepackage{subfigure}

\begin{document}

\title{Auxiliary-cavity-assisted ground-state cooling of optically levitated nanosphere in the unresolved-sideband regime}

\author{Jin-Shan Feng}
\affiliation{Institute of Theoretical Physics, Lanzhou University, Lanzhou $730000$, China}

\author{Lei Tan}
\email{tanlei@lzu.edu.cn}
\affiliation{Institute of Theoretical Physics, Lanzhou University, Lanzhou $730000$, China}

\author{Huai-Qiang Gu}
\affiliation{School of Nuclear Science and Technology, Lanzhou University, Lanzhou $730000$, China}

\author{Wu-Ming Liu}
\affiliation{Beijing National Laboratory for Condensed Matter Physics, Institute of Physics, Chinese Academy of Sciences, Beijing 100190,China}

\date{\today}
\begin{abstract}
We theoretically analyse the ground-state cooling of optically levitated nanosphere in unresolved-sideband regime by introducing a coupled high-quality-factor cavity. On account of the quantum interference stemming from the presence of the coupled cavity, the spectral density of the optical force exerting on the nanosphere gets changed and then the symmetry between the heating and the cooling processes is broken. Through adjusting the detuning of strong-dissipative cavity mode, one obtains an enhanced net cooling rate for the nanosphere. It is illustrated that the ground state cooling can be realized in the unresolved sideband regime even if the effective optomechanical coupling is weaker than the frequency of the nanosphere, which can be understood by the picture that the effective interplay of the nanosphere and the auxiliary cavity mode brings the system back to an effective resolved regime. Besides, the coupled cavity refines the dynamical stability of the system.
\end{abstract}



\maketitle


\section{Introduction}

Cavity optomechanics, providing the effective coupling between light and mesoscale matter, has been of interest in theoretical and experimental investigations\cite{Wilson,Marquardt,Kippenberg,Kipp,F.Marquardt,Aspe}. As an implementation of cavity-optomechanics, optically levitated nanosphere\cite{Zoller,Isart,Monteiro,Yin,Neukirch} in cavity is an important platform for realizing the quantum behavior at macroscale and exploring new applications of this field. The lack of the mechanical support in such levitated system leads to high mechanical quality factor and long coherence time. These benefits make the system have prominent advantages in ultrasensitive measurement\cite{Vernooy,Libbrecht,Geraci,Arvanitaki,LiTZ2,Millen,Moore,Ranjit,LIJ,Ranjit1,Rider,LiuJ,Aranas} and tests of fundamental theories that includes Nonlinear\cite{Xuereb,Gieseler0,Gieseler,Genoni,Ge,Fonseca,Rashid,XIAO}, Nonequilibrium\cite{Gieseler15,Gieseler1}, Macroscopic quantum behavior\cite{Isart1,Isart2,Bateman,Zhang,YIN}, and so on\cite{LITZ1,Nie3,LiTZ3,Nie2,Abdi,Goldwater,Zabolotskii,Honang,LIU,Minowa}. Although remarkable advances have been seen for the levitated nanosphere system, many related studies and highly sensitive measurements are still limited by the thermal noise. So it is a prior condition for all work to cool the nanosphere\cite{Barker,TZLi,Pender,Nie,Arita,Mestres,Millen1,Rodenburg,Frimmer,Jain} as micromechanical resonator all the way to quantum ground-state.

The cooling utilizing radiation pressure of levitated nanosphere in a cavity\cite{Yin1,Gieseler2,Asenbaum,Kiesel} is based on the principle that the scattering process related to cooling (anti-Stokes process) can be enhanced by choosing appropriate detuning between driving field and the cavity mode\cite{Teufel}. This requires the levitated nanosphere system to be in the ``resolved-sideband" regime, where the cavity linewidth should be smaller than the mechanical oscillator resonance frequency. Such requirement is stringent for the levitated nanosphere system characterized by low oscillation frequency ($<1$ MHz) with large cavity-decay. On the other hand, low-frequency nanosphere has large zero-point motion, and methods for cooling such nanosphere to the quantum regime are beneficial to new technical applications as well as fundamental studies. Therefore, extending the cooling domain to the unresolved-sideband regime has favorable prospect for cavity optomechanics. There have been several specific proposals, such as the dissipative coupling mechanism\cite{Elste,LiM,Xuereb1,Weiss,Yan}, parameter modulations\cite{LiY,Liao,Wang,Machnes}, hybrid system approaches\cite{Genes,Hammerer,Purdy,Paternostro,GENES,Camerer,Vogell,Gu,Bariani,Dantan,Bennett,Restrepo,Ojanen,Guo,Ranjit2,Chen,Nie1,Sarma,Zhou}, etc.\cite{LIUYC,Asjad,Yasir}, to achieve the ground-state cooling in unresolved-sideband regime\cite{LiuYC} for the cavity optomechanical system. However, these proposals are still hard to effortlessly realize. A pragmatic scenario to relax the limitation of resolved-sideband is to enhance the effective optomechanical response of the nanosphere, by coupling the cavity to an auxiliary quantum system which can be easily prepared in the experiment\cite{Liu}.

In this article, we couple the large damping optomechanical cavity in the levitated nanosphere system to an additional high-Q cavity. The large decay rate of optomechanical cavity in this system retains the efficiency for cooling through the interaction between the high-Q cavity and the optomechanical cavity. Due to no direct coupling between the auxiliary cavity and the nanosphere, the parameters of the optical and the mechanical properties can be optimized individually. We show the destructive quantum interference behavior in optical force spectrum changes the symmetry between the cooling and the heating processes of the nanosphere. One can obtain a high cooling rate at large optomechanical cavity decay rate by tuning the optical parameters of the two cavities. The cooling rate for the coupled-cavity system has  two abnormal phenomena: different from the red detuning for optimum cooling in single-cavity system, the maximum cooling rate in the coupled-cavity system corresponds to blue tuning; while the large cooling rate  in the single-cavity system is available only at low optomechanical cavity decay rate, it is achievable for both small and large optomechanical cavity damping in the case of coupled-cavity system. For our model, the ground-state cooling can be achieved under the condition that the effective coupling between the cooling field and the nanosphere is weaker than the frequency of the nanosphere. To comprehend this result and explain the abnormal phenomena in cooling process, we derive an effective indirect coupling between the auxiliary cavity and the nanosphere, which refines the dynamical stability of the system compared with the case without auxiliary cavity.

The paper is organized as follows. In Sec. \ref{1} we describe the Hamiltonian of the system and in Section \ref{2} derive the quantum Langevin equations for the system operators. Section \ref{3} is devoted to the analysis of cooling of the nanosphere. The coherence coupling and the dynamical stability of this system will be discussed in Sec. \ref{4}, followed by the conclusion of our work in Sec. \ref{5}.

\section{Model Hamiltonian}\label{1}
\begin{figure}[ht]
\begin{center}
\includegraphics[width=0.8\textwidth]{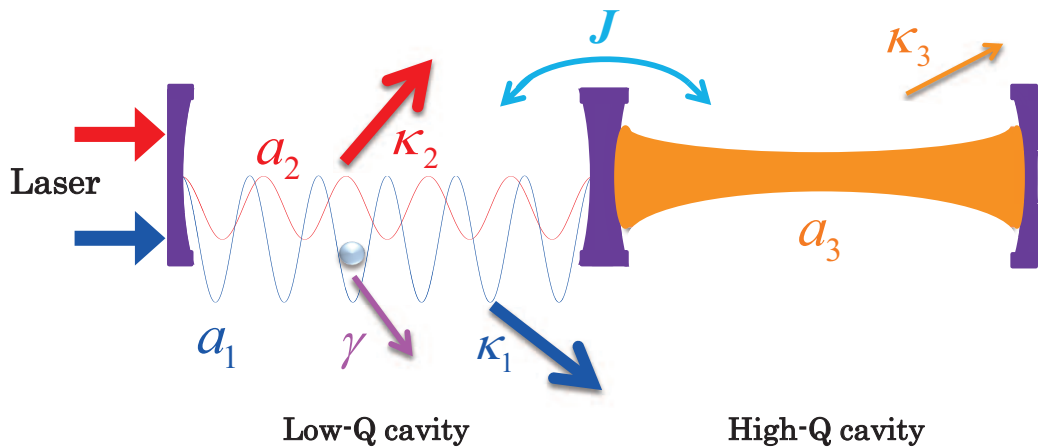}
\caption{(Color online) Hybrid optomechanical setup containing two coupled optical cavities. The first cavity with an optically levitated nanosphere has a low-Q, while the second one has a high-Q and doesn't interact with the levitated nanosphere. These two cavities are coupled by the tunneling optical mode.}
\label{Fig:1}
\end{center}
\end{figure}
The system we consider includes two coupled cavities, as shown in Fig. \ref{Fig:1}. The first one provides a simple optomechanical system, of which the mechanical part is formed by an optically levitated nanosphere\cite{Zoller}. The dielectric nanosphere is manipulated by two spatial modes of this cavity. It is confined to an optical dipole trap \cite{Grimm} provided by one mode (denoted mode 1) which is driven resonantly. The other mode (denoted mode 2) driven by a weaker beam provides a radiation pressure for cooling the motion of the nanosphere. The second cavity supports an auxiliary field (denoted mode 3), which does not interact with the nanosphere directly. The coupling between two cavities is realized by the hopping through the joint mirror of photons in them\cite{Grudinin,Grudinin1,Zheng,Peng,Xu,Xiao,Sato,Cho,LiBB}. The joint mirror has no contribution for the decay of each cavity. The dissipative nature of cavities is determined by the mirrors at two side. They are driven by corresponding pump laser. In what follows, we refer to them as optomechanical cavity and auxiliary cavity, respectively. The Hamiltonian of the system is given in a rotating frame (with $\hbar=1$) by\cite{Zoller,Liu}
\begin{eqnarray}\label{Hamiltonian}
\hat H &=&  - {\Delta _1}\hat a_1^ \dag {{\hat a}_1} - {\Delta_2}\hat a_2^\dag {{\hat a}_2} - {\Delta_3}\hat a_3^ \dag {{\hat a}_3}+ \frac{{\hat p^2}}{{2m}} \nonumber \\
&&-g_1\hat a_1^ \dag {{\hat a}_1}({\cos2k_1}{\hat x}-1) - g_2\hat a_2^ \dag {{\hat a}_2}{\cos 2(k_2{\hat x}-\frac{\pi}{4})} + J{\hat a_2^ \dag {{\hat a}_3}} + J^\ast{\hat a_3^ \dag {{\hat a}_2}} \nonumber \\
 && +({E_1^\ast}\hat a_1 + {E_1}{\hat a_1^\dag}) + ({E_2^\ast}\hat a_2 + {E_2}{\hat a_2^\dag})+({E_3^\ast}\hat a_3 + {E_3}{\hat a_3^\dag}).
\end{eqnarray}
The first line represents the free Hamiltonian of the system, where $\Delta_1 = \omega_1- \omega_o $, $\Delta_2 = \omega_2- \omega_o$ and $\Delta_3= \omega_3- \omega_a$ are the detunings between the driving field and cavity mode frequencies. $\omega_{i~(i=1,2,3)}$, $\omega_o$, and $\omega_a$ correspond to the pump fields, optomechanical cavity mode and auxiliary cavity mode frequencies, respectively. $\hat a_{i~(i=1,2,3)}$ is the annihilation operator for the corresponding cavity mode, $\hat p$ is the momentum operator of the center-of-mass of the nanosphere, and $m$ is the mass of the nanosphere.

The interactions are described by the second line. The previous two terms $g_1\hat a_1^ \dag {{\hat a}_1}({\cos2k_1}{\hat x}-1)$ and $g_2\hat a_2^ \dag {{\hat a}_2}{\cos 2(k_2{\hat x}-\frac{\pi}{4})}$ correspond to the optomechanical coupling of optical modes $\hat a_{1,2}$ with the nanosphere. $g_{i~(i=1,2)} = \frac{3V}{4V_{c,i}} \frac{\epsilon - 1}{\epsilon + 2} \omega_i$ quantifies the optomechanical interaction strength, where $V$ and $V_{c,i}$ are the nanosphere and the corresponding optical modes volumes, $\epsilon$ is the dielectric constant of the nanosphere, and $\hat x$ is the position operator of the nanosphere\cite{Zoller}. The two remaining terms $J{\hat a_2^ \dag {{\hat a}_3}}$ and $ J^\ast{\hat a_3^ \dag {{\hat a}_2}}$ stand for the interplay between optomechanical and auxiliary cavity modes. The tunnel-coupling strength of the cavities is characterized by the parameter $J$. This parameter is more difficult to define precisely because of the more detail technical factors, such as the material property of the joint mirror, the mode matching, etc. Phenomenologically, we neglect the losses in hopping and assume the mode matching is perfect. Applying Input-Output Relations, we set the $J = \sqrt{\kappa_{2} \kappa_{3}}$\cite{Bariani,Gardiner}.

 The last line accounts for the optical driving, with $E_1 = \sqrt{\kappa_1^{ex}P_1/\hbar\omega_1}e^{i\phi_1}$, $E_2 = \sqrt{\kappa_2^{ex}P_2/\hbar\omega_2}e^{i\phi_2}$ and $E_3 = \sqrt{\kappa_3^{ex}P_3/\hbar\omega_3}e^{i\phi_3}$ the amplitudes of pump lasers, $P_{i~(i=1,2,3)}$ the input powers, $\kappa_{i~(i=1,2,3)}^{ex}$  the decay rates of the photons into the associated outgoing mode and $\phi_{i~(i = 1,2,3)}$ the initial phases for the input lasers\cite{Liu}.

Based on the fact that $\omega_1,\omega_2 \gg |\omega_1 - \omega_2|$, we assum that mode $1$ and $3$ have semblable poperties, so $\omega_1 \approx \omega_2 = \omega, k_1 \approx k_2 = k, g_1 \approx g_2= g, \kappa_1 = \kappa_2 = \kappa$, for simplicity.

\section{ Heisenberg Motion Equation and Linearization}\label{2}

From the Hamiltonian given by Eq. (\ref{Hamiltonian}), we obtain the Heisenberg-Langevin equations of the system operators:
\begin{eqnarray}
&&\dot{\hat{a}}_1 =( i{\Delta _1}-\frac{\kappa}{2}){\hat a}_1 - iE_1 + \sqrt{\kappa}{\hat a_{in,1}},\nonumber\\
&&\dot{\hat{a}}_2 =[ i({\Delta _2} + 2gk\hat x) - \frac{\kappa}{2}]{{\hat a}_2} - iJ{\hat a_3} - iE_2 + \sqrt{\kappa}{\hat a_{in,2}}, \nonumber \\
&&\dot{\hat{a}}_3 =( i{\Delta _3}-\frac{\kappa_3}{2}){\hat a}_3 - i{J^\ast}{\hat a_2} - iE_3 + \sqrt{\kappa_3}{\hat a_{in,3}}, \nonumber \\
&&\dot{\hat{p}} = - 4g{k^2}{\hat a_1^ \dag {{\hat a}_1}}{\hat x} + 2gk\hat a_2^\dag {{\hat a}_2} - \frac{\gamma}{2}\hat p + \hat F_p(t),\nonumber \\
&&\dot{\hat{x}} = \frac{\hat p}{m},
\end{eqnarray}
where $\kappa$ and $\kappa_3$ are the cavity mode loss of optomechanical and auxiliary cavities, respectively. $\gamma$ is the dissipation rate of the nanosphere motion. $\hat a_{in, 1}$, $\hat a_{in, 2}$, and $\hat a_{in, 3}$ are the input vacuum noise operators, which have zero mean values and obey the nonzero correlation functions given $\langle\hat a_{in, 1}(t)\hat a^\dag_{in, 1}(t')\rangle = \langle\hat a_{in, 2}(t)\hat a^\dag_{in, 2}(t')\rangle = \langle\hat a_{in, 3}(t)\hat a^\dag_{in, 3}(t')\rangle = \delta( t - t')$, $\langle\hat a^\dag_{in, 1}(t)\hat a_{in, 1}(t')\rangle = \langle\hat a^\dag_{in, 2}(t)\hat a_{in, 2}(t')\rangle = \langle\hat a^\dag_{in, 3}(t)\hat a_{in, 3}(t')\rangle = 0$\cite{Gardiner}. $\hat F_p(t)$ is the damping force acting on the sphere with zero mean value, which obeys the following correlation function\cite{Giovannetti}:
\begin{equation}
  \langle\hat F_p(t)\hat F_p(t')\rangle = \frac{\hbar \gamma m}{2\pi}\int d\omega e^{-i\omega(t-t')} \omega \left[1 + \coth(\frac{\hbar\omega}{2k_BT})\right].
\end{equation}
Here $k_B$ is the Boltzmann constant and $T$ is the thermal bath temperature related to the nanosphere.

\begin{figure} [ht]
  \begin{center}
  \includegraphics[width=0.4\textwidth]{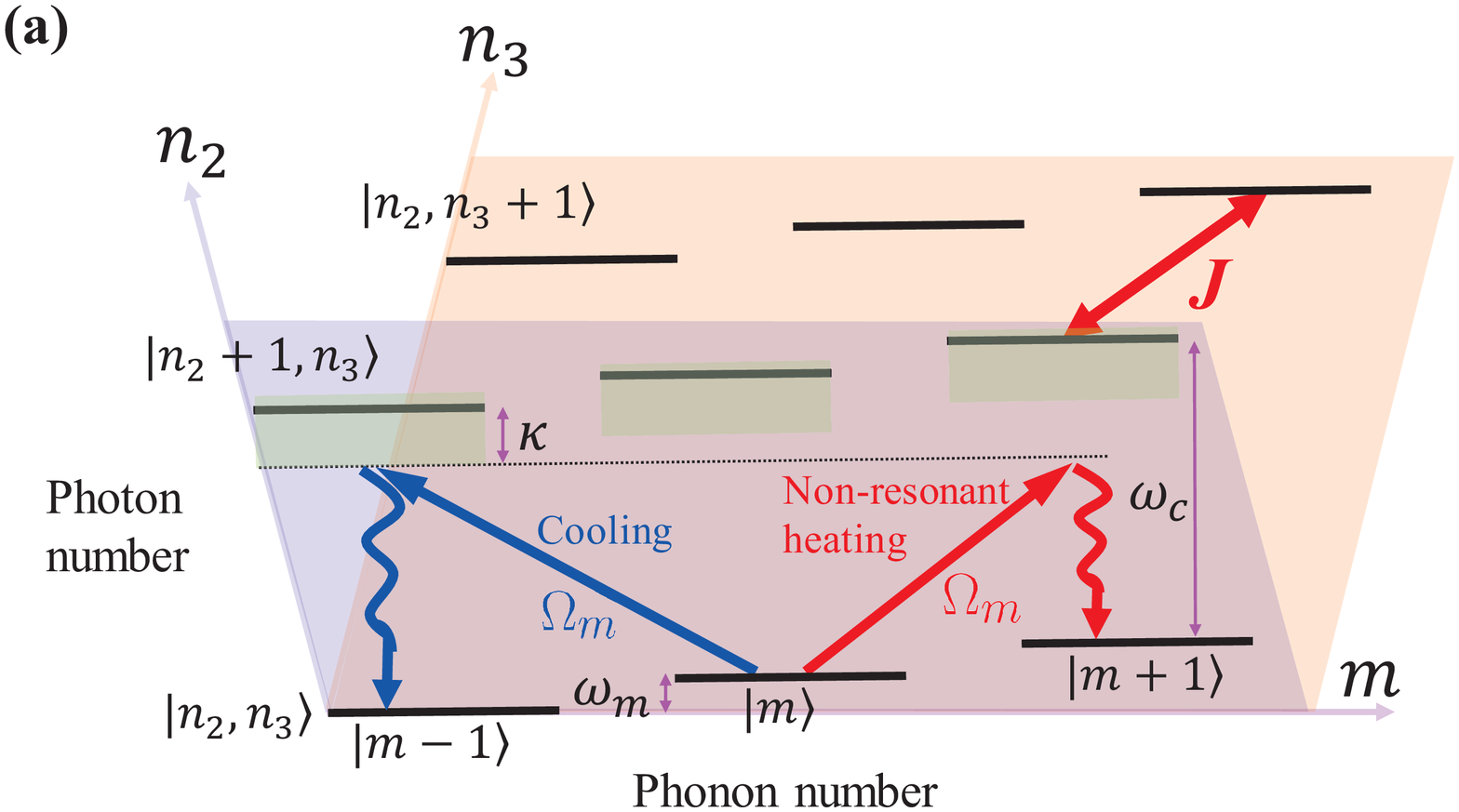}
  ~~~~
  \includegraphics[width=0.4\textwidth]{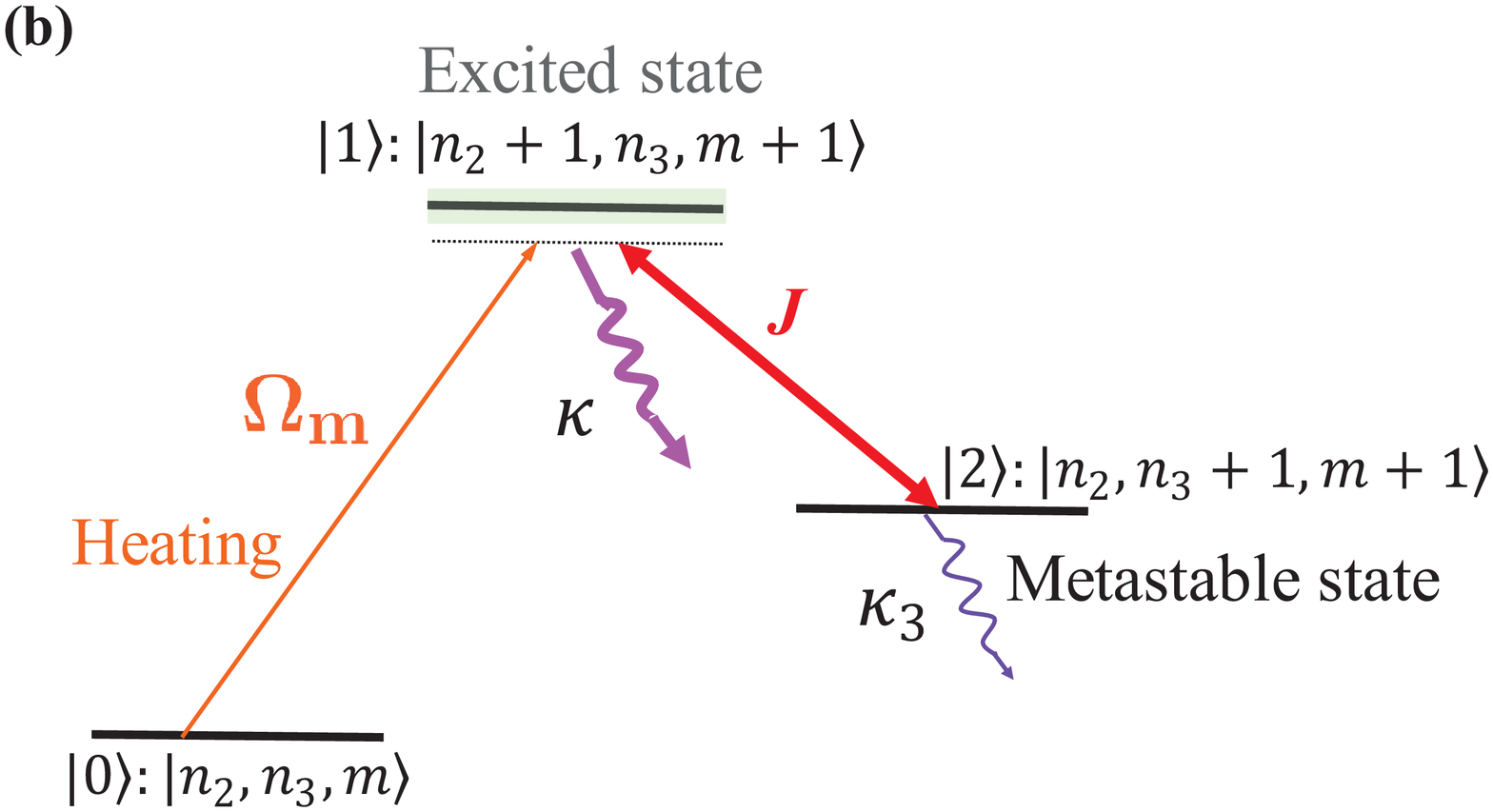}
  \caption{(Color online) (a) Energy level diagram of the linearized Hamiltonian (see Eq. (\ref{LHamiltonian})). Here $|n_2, n_3, m\rangle$ denotes the state for $n_2$ number cooling field photons in optomechanical cavity, $n_3$ number photons in auxiliary cavity, and $m$ number phonons in mechnical mode of the nanosphere. The one-way arrows represent the cooling (blue arrows) and heating (red arrows) processes due to sideband resonance. The transition between energy levels of two coupled cavities is denoted by red double arrow. (b) The three-level configuration extracted from the Fig. \ref{level} (a). State $|1\rangle$ stands for a short-lived state with high decay rate $\kappa$, and $|2\rangle$ represents a long-lived metastable state with small decay rate $\kappa_3$. It should be pointed out emphatically that the levels $|n_2 + 1, n_3\rangle$ and $|n_2 , n_3 + 1\rangle$ have obvious interval in this figure, but they are degenerate.}
  \label{level}
  \end{center}
\end{figure}

Under the condition of strong driving, we can linearize the Eqs. (3.1) around the steady-state mean values by using the transformation $\hat{a}_1 \rightarrow \alpha_1 + a_1 $, $\hat{a}_2 \rightarrow \alpha_2 + a_2 $, $\hat{a}_3 \rightarrow \alpha_3 + a_3 $, $\hat{x} \rightarrow x_0 + x$, where $\alpha_1, \alpha_2, \alpha_3$ and $x_0$ are the mean values of the operators, and $a_1, a_2, a_3$ and $x$ are the small fluctuating terms. After segregating the mean values and the fluctuating terms, we obtain the equations for the steady-state expectation values of the nanosphere and cavity field
 \begin{eqnarray}
 &&0 = -\frac {\kappa}{2}{\alpha _1}- iE_1, \\
 &&0 =[ i({\Delta _2} + 2gk{x_0}) - \frac{\kappa}{2}]{\alpha _2} - iJ{\alpha _3} - iE_2 ,\\
 &&0 =( i{\Delta _3}  - \frac{\kappa_3}{2}){\alpha _3} - i{J^\ast}{\alpha _2} - iE_3 , \\
 &&0 = -4g{k^2}{\left| {{\alpha _1}} \right|^2}x_0 + 2gk{\left| {{\alpha _2}} \right|^2},\\
 &&0 = p_0.
 \end{eqnarray}
By neglecting the higher-order terms and choosing $\Delta_1 =0$, the linear quantum Langevin equations read
 \begin{eqnarray}
 {\dot a_1} &=& -i4g{k^2}{x_0}{\alpha _1}x - \frac{\kappa}{2}{a_1} + \sqrt{\kappa}{\hat a_{in,1}} , \\
 {\dot a_2} &=& [ i({\Delta _2} + 2gk{x_0}) - \frac{\kappa}{2}]{a _2} + 2ig{\alpha _2}kx - iJ{a_3} + \sqrt{\kappa}{\hat a_{in,2}} , \\
 {\dot a_3} &=& ( i{\Delta _3} - \frac{\kappa_3}{2}){a _3} - i{J^\ast}{a_2} + \sqrt{\kappa_3}{\hat a_{in,3}} , \\
 {\dot p} &=& - 4g{k^2}{\left| {{\alpha _1}} \right|^2}x - \frac{\gamma}{2} p + \hat F_p(t)
            + 2gk[{\alpha_2}{a_2^\dag} + {\alpha_2^\ast}{a_2} - 2k{x_0}({\alpha_1}{a_1^\dag} + {\alpha_1^\ast}{a_1})],\\
 {\dot x} &=& \frac{p}{m}.
 \end{eqnarray}
 From Eq. (3.11), we note that cavity mode 1 provides a linear restoring force $dp/dt \sim - 4g{k^2}{\left| {{\alpha _1}} \right|^2}x = -m\omega^2_mx$. $\omega_m$ is the harmonic oscillator frequency of the nanosphere. The corresponding linearized system Hamilton is written as
\begin{eqnarray}\label{LHamiltonian}
 H &=&  - {\Delta_1} a_1^ \dag {a_1} - {\Delta'_2} a_2^\dag {a_2} - {\Delta_3} a_3^ \dag {a_3} + \frac{p^2}{2m} + 4gk^2|\alpha_1|^2x^2 \nonumber \\
&&-(\Omega_ma_2^\dag +\Omega_m^\ast a_2)(b^\dag + b) + J{ a_2^ \dag { a_3}} + J^\ast{a_3^ \dag {a_2}},
\end{eqnarray}
where $\Delta'_2 = \Delta _2 + 2gkx_0$ is the detuning relative to the new resonance frequency of the optomechanical cavity, $b = \frac{x}{x_m} + i\frac{p}{\sqrt{2m\hbar\omega_m}}$ is the annihilation operator of the mechanical mode. $\Omega_m = 2gkx_{ZPF}\alpha_2$ is the effective optomechanical coupling strength and $x_{ZPF} \equiv \sqrt{\hbar/2m\omega_m}$ is the zero-point fluctuation of the nanosphere. The energy levels for the linearized Hamiltonian are demonstrated in Fig. \ref{level} (a). The transition processes among levels contain two parts. The primary one is the cooling and heating processes on account of the interaction between the cooling optical mode and the nanosphere in optomechanical cavity\cite{Zoller,Liu1}, which are denoted by the one-way arrows. the other is the energy swapping of the optomechanical and the auxiliary cavities due to the tunneling between them, which is labeled by the red double arrows. It is worthwhile to mention that the energy level structure of the system is transformed from a two-level to a three-level after the auxiliary cavity is added, as in Fig. \ref{level} (b).

\section{Cooling of Nanosphere}\label{3}

\subsection{Optical force spectrum}
From the interaction term $-(\Omega_ma_2^\dag +\Omega_m^\ast a_2)(b^\dag + b)$ in Eq. (3.13), we can derive the optical force on the nanosphere $F = (\Omega_ma_2^\dag +\Omega_m^\ast a_2)/x_{ZPF}$. By the the Fourier transformation of the correlation function, the corresponding quantum noise spectrum is expressed as $S_{FF}(\omega) \equiv \int \langle F(t)F(0)\rangle e^{i\omega t}dt$. To gain the analytic expression of this noise spectrum, we treat the optomechanical coupling as a perturbation to the optical field because of the strong dissipative nature of optomechanical cavity. Firstly, we transform the corresponding linear motion equations to the frequency domain, i.e.
\begin{eqnarray}
 -i\omega \tilde{a}_2(\omega) &=& ( i{\Delta' _2}  - \frac{\kappa}{2})\tilde{a }_1(\omega) + i\Omega_m [\tilde{b}^\dag(\omega) + \tilde{b}(\omega)]- iJ\tilde{a}_3(\omega) + \sqrt{\kappa}\tilde{a}_{in,2}(\omega), \\
 -i\omega \tilde{a}_3(\omega) &=& ( i{\Delta _3} - \frac{\kappa_3}{2})\tilde{a }_3(\omega) - i{J^\ast}\tilde{a}_2(\omega) + \sqrt{\kappa_3}\tilde{a}_{in,3}(\omega), \\
  -i\omega \tilde{b}(\omega) &=& (- i\omega_{m} - \frac{\gamma}{2}) \tilde{b}(\omega) + i[\Omega_m\tilde{b}^\dag(\omega) + \Omega_m^\ast\tilde{b}(\omega)] + \sqrt{\gamma}\tilde{b}_{in}(\omega).
\end{eqnarray}
Then we derive the expression for $\tilde{b}(\omega)$ as
\begin{equation}
  \tilde{b}(\omega) \simeq \frac{\sqrt{\gamma}\tilde{b}_{in}(\omega) + i\sqrt{\kappa}A_2(\omega) + \sqrt{\kappa_3}A_3(\omega)}{i\omega - i[\omega_m + \Sigma(\omega)] - \frac{\gamma}{2}},
\end{equation}
where\begin{eqnarray}
       A_2(\omega) &=& \Omega^\ast_m \chi(\omega)\tilde{a}_{in,2}(\omega) + \Omega_m\chi^\ast(-\omega)\tilde{a}^\dag_{in,2}(\omega),\\
       A_3(\omega)&=& J[\Omega^\ast\chi(\omega)\chi_3(\omega)\tilde{a}_{in,3}(\omega) - \Omega\chi^\ast(-\omega)\chi_3^\ast(-\omega)\tilde{a}^\dag_{in,3}(\omega)], \\
       \Sigma(\omega) &=& -i|\Omega_m|^2[\chi(\omega) - \chi^\ast(\omega)],\\
       \chi_2(\omega) &=& \frac {1}{-i(\omega + {\Delta _2^\prime}) + {\kappa}/{2}},\\
       \chi_3(\omega) &=& \frac {1}{-i(\omega + {\Delta _3}) + {\kappa_3}/{2}},\\
       \chi(\omega) &=& \frac {1}{\frac{1}{\chi_{2}(\omega)} + \left| J \right|^2{\chi_{3}(\omega)}},\\
       \chi_m(\omega) &=& \frac {1}{-i(\omega - \omega_{m}) + {\gamma}/{2}}.
     \end{eqnarray}
Here the effect of the optomechanical and the auxiliary cavities is represented by $A_{2, 3}(\omega)$. $\Sigma(\omega)$ accounts the optomechanical self-energy; $\chi(\omega)$ is the total response function of two coupled cavities, and $\chi_2(\omega),\chi_3(\omega),$ and $\chi_m(\omega)$ are the response function of the optomechanical cavity, the auxiliary cavity, and the mechanical mode, respectively. The influence of the optomechanical coupling on the nanosphere motion is the modification of its mechanical frequency $\delta\omega_m = Re\Sigma(\omega_m)$ and damping $\Gamma_{opt} = -2Im\Sigma(\omega_m)$.

With the above prepation, we obtain the spectral density for the optical force:
\begin{eqnarray}\label{spectrum}
  S_{FF}(\omega) &=& \frac{|\Omega_m\chi(\omega)|^2}{x^2_{ZPF}}\left[\kappa + \kappa_3|J|^2|\chi_3(\omega)|^2\right]\nonumber \\
   &=& \frac{|\Omega_m |^2}{x^2_{ZPF}}\left|\frac{1}{-i(\Delta'_2 + \omega) + \frac{\kappa}{2} + \frac{|J|^2}{-i(\Delta_3 + \omega) + \frac{\kappa_3}{2}}} \right|^2\left(\kappa + \frac{\kappa_3 |J|^2}{(\Delta_3 + \omega)^2 + \frac{\kappa_3^2}{4}}\right).
\end{eqnarray}

\begin{figure*}[!htb]
\centering
\includegraphics[width=0.403\linewidth]{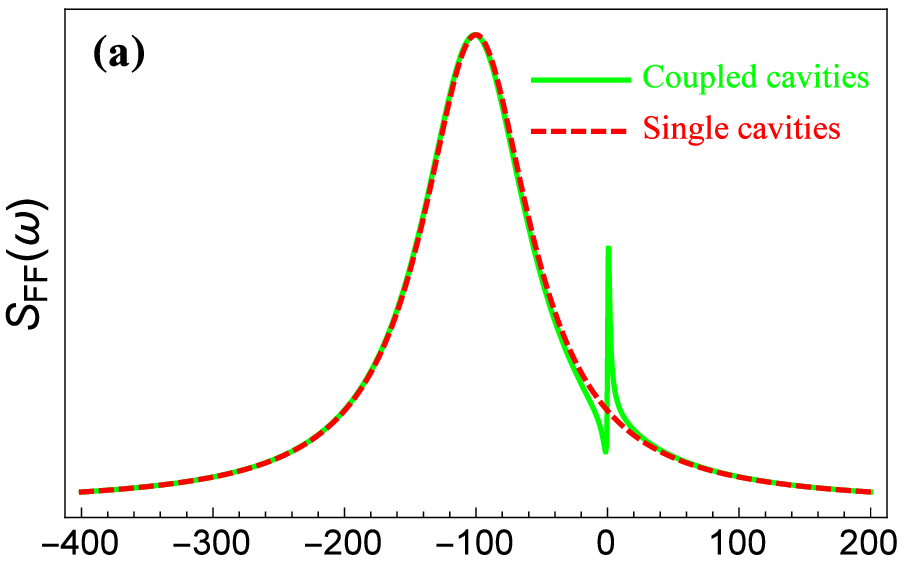}
\includegraphics[width=0.4\linewidth]{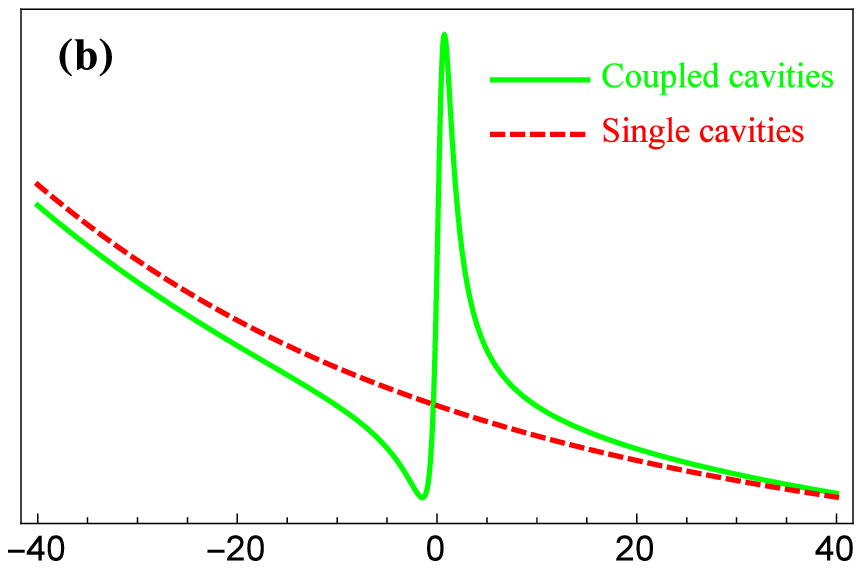}\\
\includegraphics[width=0.403\linewidth]{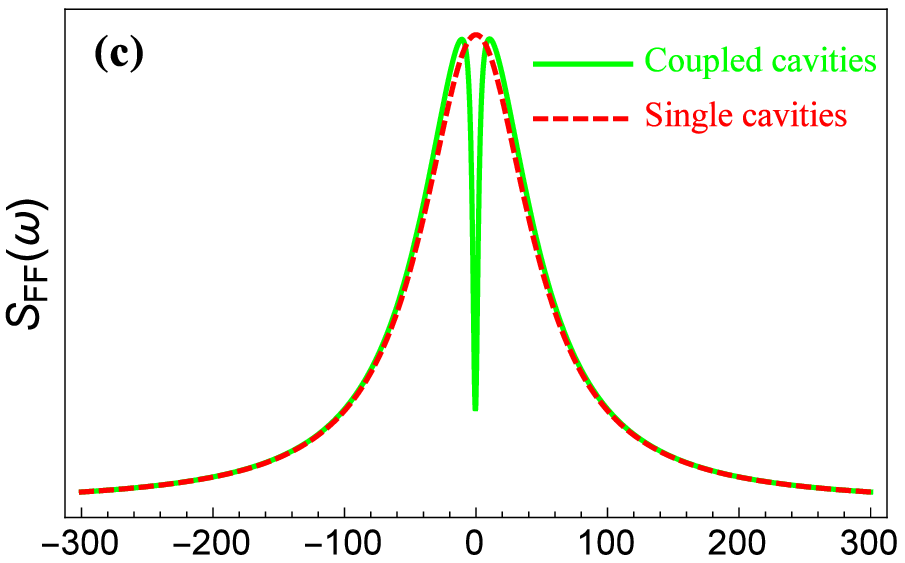}
\includegraphics[width=0.4\linewidth]{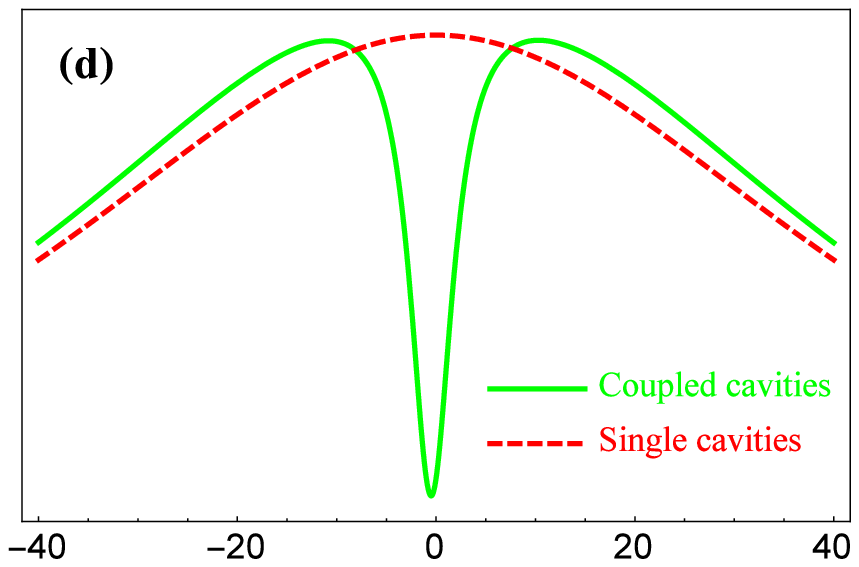}\\
\includegraphics[width=0.403\linewidth]{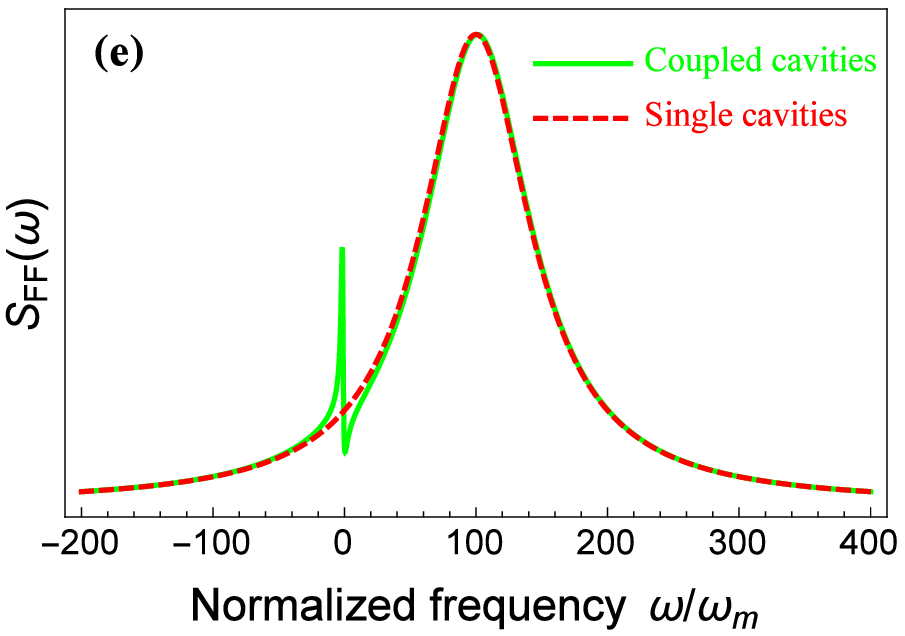}
\includegraphics[width=0.4\linewidth]{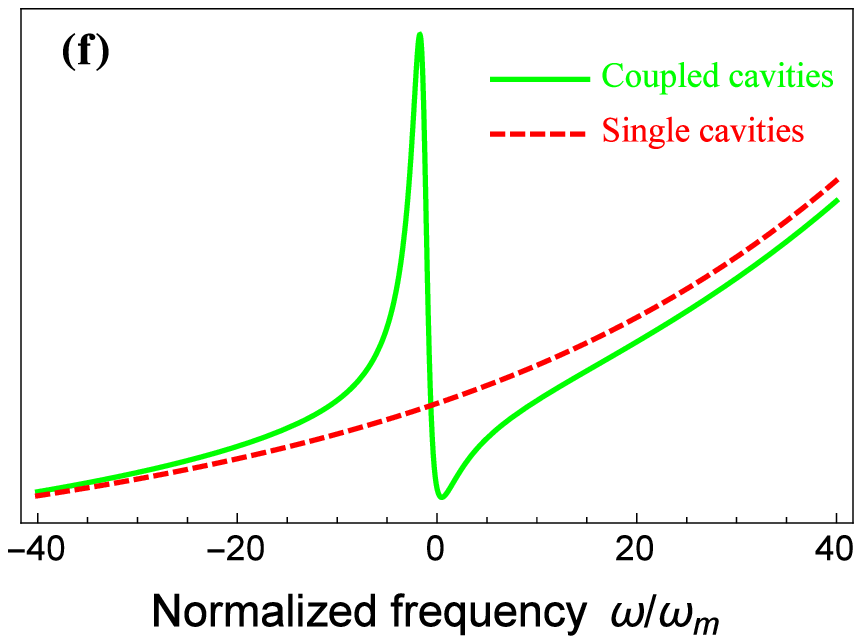}
\caption{(Colour online) Optical force spectrum $S_{FF}(\omega)$ of single cavity and coupled cavities vs normalized frequency $\omega / \omega_m$ for various normalized detuning $\Delta'_2/\omega_m$.
(a) The spectrum for blue detuning $\Delta'_2 = 100\omega_m$.
(b) Detail view of (a) for Fano line shape.
(c) The spectrum for resonant $\Delta'_2 = 0$.
(d) Detail view of (c) for EIT-like line shape.
(e) The spectrum for red detuning $\Delta'_2 = -100\omega_m$.
(f) Detail view of (e) for Fano line shape.
The other parameters are $\Delta_3 = 0.5\omega_m, \kappa/\omega_m = 100, \kappa_3 = \omega_m, J = \sqrt{\kappa\omega_m}, \Omega_m = 5\omega$, and $\gamma = 10^{-5}\omega_m$.}
\label{spectrumplot}
\end{figure*}

For a general cavity optomechanical system, the noise spectrum has the form of $S_{FF}(\omega) = |\Omega_m\chi(\omega)|^2\kappa/x^2_{ZPF}$, which equals to Eq. (\ref{spectrum}) choosing $J=0$. This is a typical Lorentzian lineshape. From Eq. (\ref{spectrum}), it can be observed that the spectral density of coupling cavities system has a complex modification comparing to single cavity. This result roots from the interaction of two optical modes when the auxiliary cavity is added. The spectral density of the optical force $S_{FF}(\omega)$ for both single cavity and coupled cavities with different type of detuning values in the unresolved-sideband are depicted in Fig. \ref{spectrumplot}. From Fig. \ref{spectrumplot} (a), (c) and (e), we find that the noise spectra of the single cavity and the coupled cavities are identical in the range far away from the resonant region of the auxiliary cavity, while a new lineshape appears in the resonant region of the auxiliary cavity for coupled cavities system. The feature of new resonance peaks in Fig. \ref{spectrumplot} (b), (d) and (f) is related to the position of the resonant regions of the optomechanical and the auxiliary cavities. When the resonant regions are separate, the lineshape of the new resonance peaks is an asymmetric Fano lineshape as shown in Fig. \ref{spectrumplot} (b) and (f). And for the overlapping case i. e. , $\Delta'_2\simeq\Delta_3$, the lineshape is a symmetric EIT-like lineshape. The emergence of new lineshape changes the symmetry of the background with symmetric Lorentzian lineshape.

The EIT-like line shape is a result of the interference between two resonant processes. The physical mechanism is shown in Fig. \ref{level} (b), where the transition processes $|0\rangle \to |1\rangle$ and $|0\rangle \to |1\rangle \to |2\rangle \to |1\rangle$ are indistinguishable, which causes the destructive quantum interference. Therefore, the excitation channel $|0\rangle \to |1\rangle$ corresponding to heating process is suppressed. Besides, the cooling process is intact for the off-resonance transition. The interference of resonant and nonresonant processes lead to the appearance of Fano line shape\cite{Elste,Stassi}. In this case, there is an enhancement for a certain process while the other is restrained\cite{Sarma}. This means that the symmetry between the heating and the cooling processes is modulated due to the presence of the auxiliary cavity.

So, it is a decent approach for a preferable cooling performance that the interference can be utilized by adjusting the optical parameter of the system to suppress the heating effect and enhance the cooling one.

\subsection{Cooling rate}

For our system, the cooling and heating rates $A_{\mp}$ are given by
\begin{equation}
  A_\mp = S_{FF}(\pm\omega_m) x^2_{ZPF}= \left|\frac{\Omega_m}{-i(\Delta'_2 \pm \omega_m) + \frac{\kappa}{2} + \frac{|J|^2}{-i(\Delta_3 \pm \omega_m) + \frac{\kappa_3}{2}}} \right|^2\left(\kappa + \frac{\kappa_3 |J|^2}{(\Delta_3 \pm \omega_m)^2 + \frac{\kappa_3^2}{4}}\right).
\end{equation}
The net cooling rate is defined as
\begin{equation}
   \Gamma_{opt} = A_- - A_+.
\end{equation}

\begin{figure*}[!htb]
\centering
\includegraphics[width=0.408\linewidth]{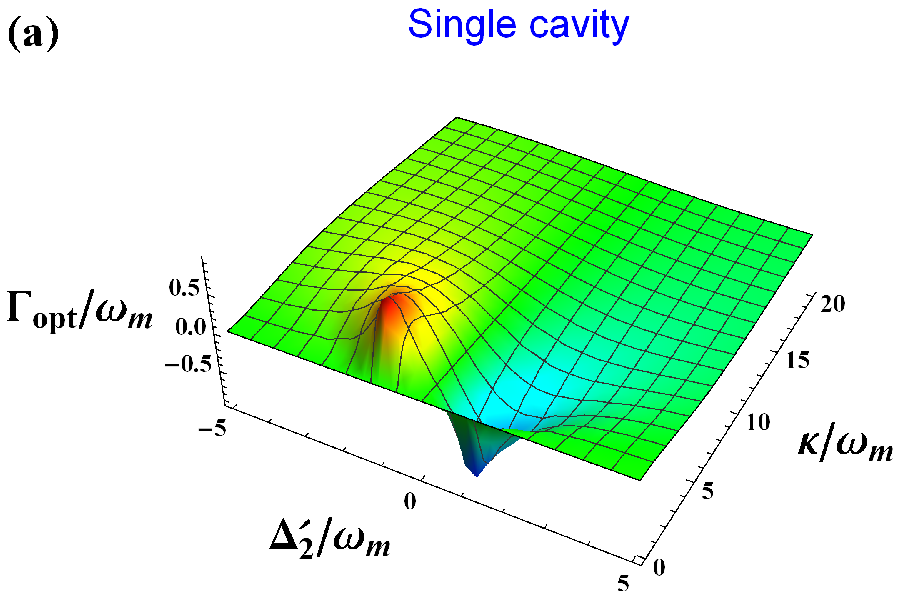}
\includegraphics[width=0.4\linewidth]{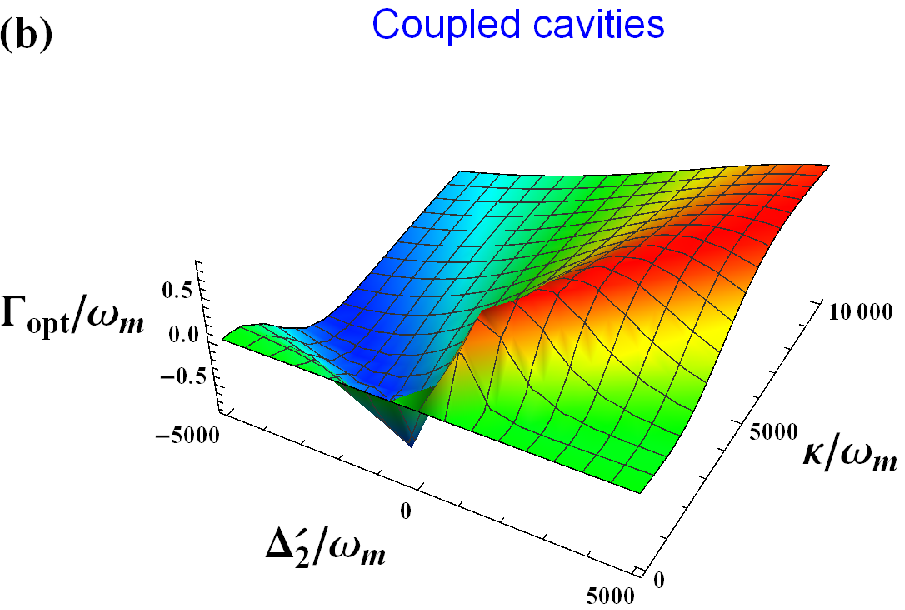}
\caption{(Color online) Net cooling rate $\Gamma_{opt}$ as functions of normalized detuning $\Delta'_2/\omega_m$ and normalized decay rate $\kappa/\omega_m$ for a single cavity (a) and coupled cavities (b). The relevant parameters are $\Delta_3 = 0.5\omega_m, \kappa_3 = \omega_m, J = \sqrt{\kappa\omega_m}, \Omega_m = \omega/4$, and $\gamma = 10^{-5}\omega_m$.}
\label{coolingrate}
\end{figure*}

In Figs. \ref{coolingrate} (a) and (b), we plot the net cooling rate for the single cavity and coupled cavities systems. There are two discrepancies between them. For the coupled cavities, the optimum cooling detuning is blue and the high cooling rate widely appears at the region of larger decay rate. Due to the existence of the auxiliary cavity, the effective detuning of the nanosphere cooling dynamics is no longer $\Delta'_2$. So $\Delta'_2 < 0$ is not the appropriate choice for the nanosphere cooling. The details will be discuss in Section \ref{discuss}. When the coupled cavities system is in the cooling regime, the cooling rate $A_-$ is unchanged while the heating rate $A_+$ is large suppression on account of the quantum interference. Consequently, a large net cooling rate is gained. The larger the damping is, the more apparently the auxiliary cavity modifies the symmetry between heating and cooling processes for extensive detuning. So a large net cooling rate for wide range is shown.

\subsection{Cooling limit}
The steady-state cooling limit (i.e. the final mean photon number) of the coupled cavities is similar to the single cavity\cite{Aspe}, which reads
\begin{eqnarray}
  n_f 
  &=& \frac{A_+}{\Gamma_{opt}}+\frac{\gamma_{sc}}{\Gamma_{opt}}.
\end{eqnarray}
The cooling limit consists of two parts. $n_f^q = A_+/\Gamma_{opt}$ is the quantum limit of cooling which relates to the quantum backaction. The classic cooling limit $n_f^c = \gamma_{sc}/\Gamma_{opt}$ is tied to the specific conditions of a particular system.

According to the above analyse, we know the quantum interference suppresses the heat rate $A_+$ in connection with the quantum backaction heating and gives rise to a larger net cooling rate $\Gamma_{opt}$. As a result, the coupled cavities system has much smaller quantum limit of cooling $n_f^q$ than the single one. With same physical quantity $\gamma_{sc}$ in both the single cavity and the coupled cavities systems, the classic cooling limit $n_f^c$ is much smaller in the coupled cavities system due to the large net cooling rate $\Gamma_{opt}$. To recap, the coupled cavities system can achieve ground-state cooling in an extensive range of parameters. In the following, a set of experimentally plausible parameters are adopted by reference to the related experiments \cite{Zoller,Pender,Rodenburg,Kiesel} to shown this result. We consider a silica sphere with radius $r = 50$ nm and mechanical frequency $\omega_m /(2\pi) = 0.5$ MHz is levitated inside a cavity with $L=1$ cm and waist $w = 25$ $\mu$m. The wavelength of the trap laser is taken $\lambda = 1$ $\mu$m and the material properties $\epsilon = 2$. Specifically, we take the effective optomechanical coupling strength $\Omega_m /\omega_{m} = 1/4 < 1$ for ensuring the validity of the perturbative result. This means the coupling between the cavity mode 2 and the nanosphere is weaker than the frequency of the nanosphere, which is different from the relevant study\cite{Sarma,Liu}. Besides, the influence origin from the background gas be negligible.

\begin{figure} [ht]
\centering
\includegraphics[width=0.4\linewidth]{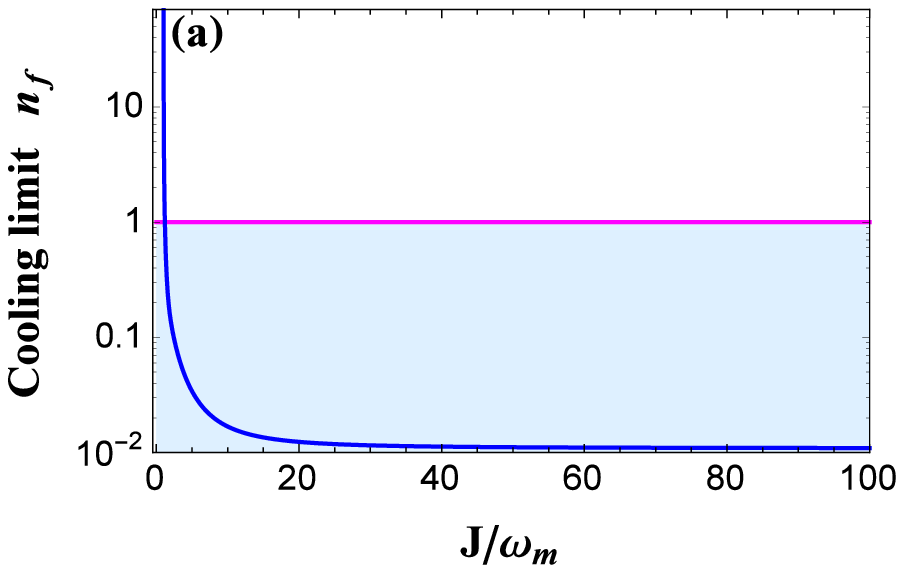}
\includegraphics[width=0.405\linewidth]{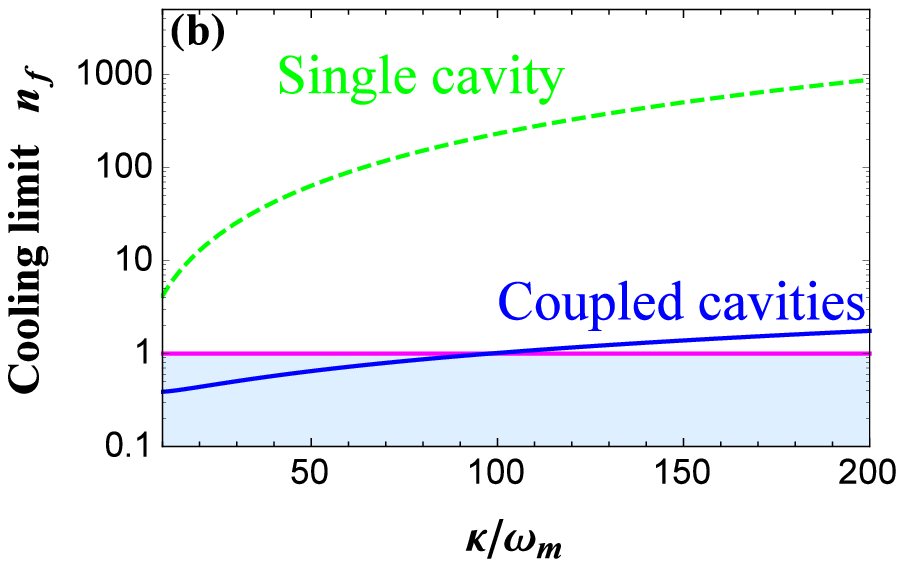}
  \caption{(Color online) The steady-state cooling limit as a function of (a) the various normalized coupling strength $J/\omega_m$ and (b) the various normalized damping $\kappa/\omega_m$ for single cavity and coupled cavities. For(a), the detuning $\Delta'_2 = \omega_m$, the decay rate $\kappa/\omega_m =(J/\omega)^2 $ . In (b), the solid blue line denotes the final phonon number of the nanosphere for couplied cavities. Meanwhile the dash green line stands for the single cavity. The shaded region denotes $n_f < 1$. The optimum detuning $\Delta'_2 = J^2/(\Delta_3 + \omega_m)$ is in accordance with Ref. \cite{Liu} and $J = \sqrt{\kappa\omega_m}$. The other parameters are $\Delta_3 = 0.5\omega_m, \kappa_3 = \omega_m,  \gamma = 10^{-5}\omega_m,$ and $\Omega_m = \omega_m/4$. For the nanosphere, the radius is chosen as $r = 50$ nm and the operating wavelength is taken $\lambda = 1$ $\mu$m.}
  \label{coolinglimit}
\end{figure}

The effect of the tunnelling strength $J$ on the cooling limit is shown in Fig. \ref{coolinglimit} (a).  We find that the ground state cooling can be achieved for a wide range of lager tunnelling strength. More carefully, the significant effect is occur at narrow region of the effective tunnelling strength increasing from zero, meanwhile the cooling limit hardly changes for enlarging the strength as the limit attains a certain value. This means the auxiliary cavity has a limit work for the nanosphere cooling. In Fig. \ref{coolinglimit} (b), we demonstrate the steady-state cooling limit of single cavity and coupled cavities for the various normalized damping $\kappa/\omega_m$. As shown in Fig. \ref{coolinglimit} (b), it is not able for the single cavity system in unresolved-sideband regime ($\kappa/\omega_m\gg1$) to cool the nanosphere to the ground-state. For the coupled cavities, due to the quantum interference originate from the addition of the auxiliary cavity, ground-state cooling can be achieved for a larger range of normalized damping $\kappa/\omega_m$.

\begin{figure*}[!htb]
\centering
\includegraphics[width=0.405\linewidth]{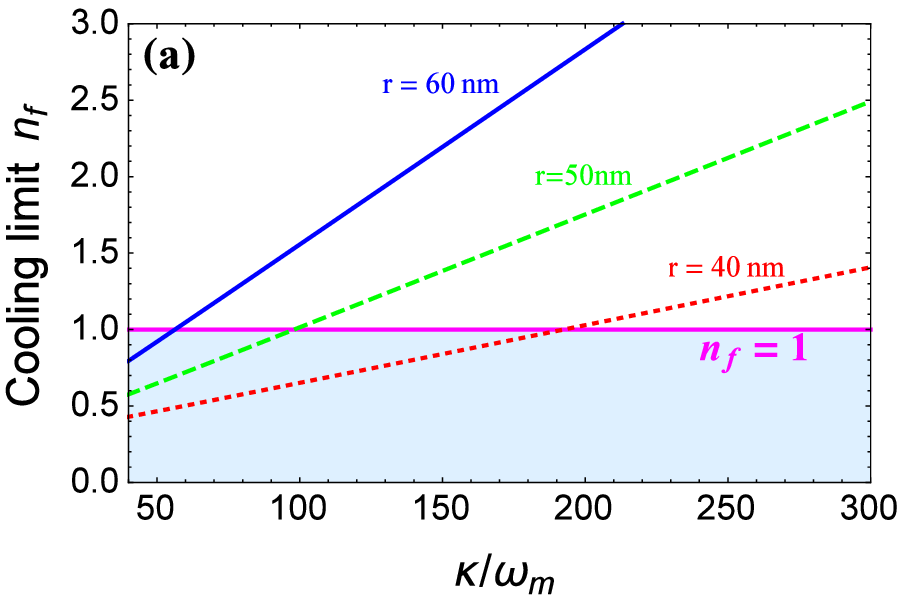}
\includegraphics[width=0.4\linewidth]{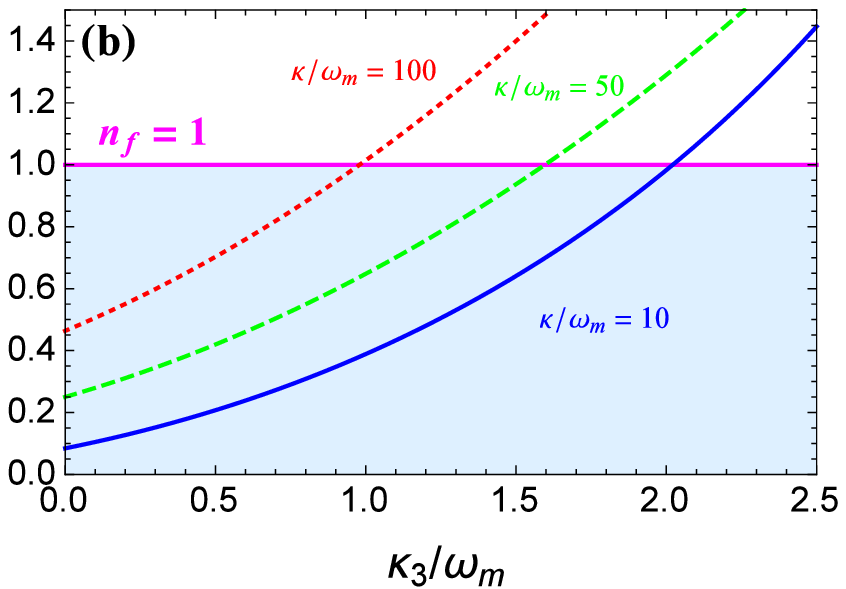}
\caption{(Color online) The cooling limit as functions of normalized damping $\kappa/\omega_m$ for different radius of the nanosphere with $\kappa_3 = \omega_m$ (a) and normalized auxiliary cavity damping $\kappa_3/\omega_m$ for different normalized decay rate $\kappa/\omega_m$ with $r = 50$ nm (b). The shaded region denotes $n_f < 1$. The relevant parameters are $\Delta_3 = 0.5\omega_m, \Delta'_2 = J^2/(\Delta_3 + \omega_m), J = \sqrt{\kappa\omega_m}, \Omega_m = \omega/4$, and $\gamma = 10^{-5}\omega_m$.}
\label{coolinglimit2}
\end{figure*}

The physical quantity $\gamma_{sc}$ in the classic cooling limit is characterized by the nanosphere volume $V$ under the condition of the same material and trap field, since $\gamma_{sc} = \omega_m\frac{4\pi^2}{5}\frac{\epsilon-1}{\epsilon+2}(V/\lambda^3)$\cite{Zoller}. The decay rate of auxiliary cavity influences the optomechanical response of the nanosphere, and then changes the cooling limit of the nanosphere. For these reasons, the radius of the nanosphere related to the nanosphere volume and the damping rate of the auxiliary cavity are crucial parameters in the coupled-cavity-nanosphere system. Fig. \ref{coolinglimit2} shows the influence of them on the cooling limit. Fig. \ref{coolinglimit2} (a) plots the cooling limit as a function of normalized damping $\kappa/\omega_m$ for different radii of the nanosphere. One finds that, the cooling limit is not sensitive to the size of the nanosphere when the decay rate $\kappa$ is small. The size of the nanosphere largely affects the cooling limit in the large decay rate $\kappa$ regime, and the nanosphere with smaller radius can achieve the ground state cooling in wide range of the parameter $\kappa$. When $\kappa$ is large, the increase of nanosphere radius will change the physical quantity $\gamma_{sc}$ rapidly, so the classic cooling limit increases too sudden to remain the nanosphere in ground state regime. Fig. \ref{coolinglimit2} (b) plots the cooling limit as a function of normalized auxiliary cavity damping $\kappa_3/\omega_m$ for different decay rate $\kappa$. It demonstrates that, the system can realize ground-state cooling in a wide range of parameter $\kappa$ under the condition $\kappa_3/\omega_m < 1$, and that the cooling limit is sensitive to decay rate $\kappa$ for $\kappa_3/\omega_m > 1$. It is more difficult to achieve ground-state cooling for larger $\kappa$ in range of $\kappa_3/\omega_m > 1$. The quantum interference leads to the actual damping of the hybrid system to relate to $\kappa_3$. When $\kappa_3/\omega_m < 1$, the hybrid system is actually in resolved regime and ground-state cooling can be obtained easily, but for $\kappa_3/\omega_m > 1$, one would obtain the opposite result (see Section \ref{discuss} for details).

\section{Discussion}\label{4}

From the above study, we know that the auxiliary cavity not only changes the symmetry between the heating and cooling processes of the nanosphere, but also modifies the cooling dynamics of the nanosphere. There exists indirect interaction between the cavity mode $a_3$ and the nanosphere. For the sake of understanding the corresponding result, we will derive the effective parameters for the coupled cavities and discuss the dynamical stability condition of our model in this section.

\subsection{Effective coupling}\label{discuss}

The current system is in the highly unresolved regime $\kappa\gg\omega_m$. The coupling between the cavity mode $a_2$ and the nanosphere is weak ($\Omega_m\ll\omega_m$), which can be taken as a perturbation. Therefore the analytical dynamical equations can be derived only for the cavity mode $a_3$ and the nanosphere. For Eqs. (3.9), (3.10) and (3.12), we derive the formal solution of the corresponding operators by formal integration:

\begin{equation}\label{e}
  a_2 = a_2(0)e^{i\Delta_{2}^{'}t-\frac{\kappa}{2}t} + e^{i\Delta_{2}^{'}t-\frac{\kappa}{2}t}\int_{0}^{t}[2ig\alpha_2kx(\tau) - iJa_3(\tau) + \sqrt{\kappa}a_{in,2}(\tau)]e^{-i\Delta_{2}^{'}\tau + \frac{\kappa}{2}\tau}d\tau,
\end{equation}

\begin{equation}\label{}
  a_3 = a_3(0)e^{i\Delta_{3}t-\frac{\kappa_3}{2}t} + e^{i\Delta_{3}t-\frac{\kappa_3}{2}t}\int_{0}^{t}[ - iJ^\ast a_2(\tau) + \sqrt{\kappa_3}a_{in,3}(\tau)]e^{-i\Delta_{3}\tau + \frac{\kappa_3}{2}\tau}d\tau,
\end{equation}

\begin{equation}\label{}
  x = \frac{p}{m}t + \int_{0}^{t}F_x(\tau)d\tau.
\end{equation}
Because $\kappa \gg J$ and $g \ll \omega_m$, we neglect the corresponding terms and obtain
\begin{eqnarray}
   a_3 &=& a_3(0)e^{i\Delta_{3}t-\frac{\kappa_3}{2}t} + A_{in,3}(t) \label{c},\\
   x &=& \frac{p}{m}t + F_X(t)\label{d},
\end{eqnarray}
where $A_{in,3}(t)$ and $F_X(t)$ represent the noise terms. Plugging Eqs. (\ref{c}) and (\ref{d}) into Eq. (\ref{e}) under the condition $|\Delta'_2|\gg|\Delta_3|, \kappa\gg(\kappa_3, \gamma)$, we have
\begin{equation}\label{f}
  a_2 = a_2(0)e^{i\Delta_{2}^{'}t - \frac{\kappa}{2}t} + \frac{2ig\alpha_2kx(t)}{-i\Delta_{2}^{'} + \frac{\kappa}{2}} - \frac{ iJ a_3(t)}{-i\Delta_{2}^{'} + \frac{\kappa}{2}} + A_{in,2}(t).
\end{equation}
Substituting Eq. (\ref{f}) into Eqs. (3.10) and (3.12) and neglecting the terms containing $e^{- \frac{\kappa}{2}t}$, one can compare the equations with the single cavity case and then derive
\begin{eqnarray}
i\Delta_3 - \frac{\kappa_3}{2} + \frac{|J|^2}{i\Delta_{2}^{'} - \frac{\kappa}{2}} &\longleftrightarrow& i\Delta_{eff} - \frac{\kappa_{eff}}{2},\\
\left|\frac{J^\ast\Omega_{m}}{i\Delta_{2}^{'} - \frac{\kappa}{2}}\right| &\longleftrightarrow& |\Omega_{m~eff}|,
\end{eqnarray}
where $|\Omega_{m~eff}| = \eta|\Omega_{m}|$, $\kappa_{eff} = \kappa_{3} + \eta^2\kappa$, $\Delta_{eff} = \Delta_{3} - \eta^2 \Delta_{2}^{'}$ and $\eta = \frac{|J|}{[\Delta_{2}^{'2} + (\frac{\kappa}{2})^2]^{\frac{1}{2}}}$.

Therefore, we reduce a three-mode system to a two-mode system\cite{Liu}. For the effective detuning $\Delta_{eff}$, because the detuning $\Delta_{3}$ is greater than zero and small as the system at cooling state, only $\Delta'_2\gg0$ (i.e. the detuning is blue) can make $\Delta_{eff} <0$ be in the optimum detuning regime. Under the condition of $\kappa\gg J$, the parameter $\eta$ is far less than $1$, so the effective decay rate $ \kappa_{eff} \simeq \kappa_{3}$. This means that the indirect coupling can brings the system from high unresolved regime to an effective resolve regime and explain why the actual damping of the hybrid system to relate to $\kappa_3$.

\subsection{Dynamical stability condition}

The dynamical stability condition of the system is derived by the Routh-Hurwitz criterion\cite{Routh}. For the single cavity system, the dynamical stability condition reads
\begin{equation}\label{stability}
  \Delta_{2}^{'}[16\Delta_{2}^{'}|\Omega_{m}|^2 + (4\Delta_{2}^{'2} + \kappa^{2} )\omega_m] < 0.
\end{equation}
When the system is in the resolved regime, the detuning for the optimum cooling limit is $\Delta_{2}^{'} = -\kappa/2$. Thus Eq. (\ref{stability}) is simplified as
\begin{equation}\label{a}
  |\Omega_{m}|^2 < \frac{\kappa\omega_{m}}{4}.
\end{equation}

The dynamical stability condition for the coupled cavities is given in terms of the derived effective parameters
\begin{equation}\label{stability1}
  \Delta_{eff}[16\Delta_{eff}|\Omega_{m~eff}|^2 + (4\Delta_{eff}^{2} + \kappa_{eff}^{2} )\omega_{m~eff}] < 0.
\end{equation}
Similarly, we take the effective detuning $\Delta_{eff} = -\omega_{m}$ which is the optimum detuning for effective optomechanical interaction in the resolved regime. Then Eq. (\ref{stability1}) reduces to
\begin{equation}
  |\Omega_{m~eff}|^2 < \omega_{m}^2/4 + \kappa_{eff}^{2}/16.
\end{equation}
Back to real parameters, we have
\begin{equation}\label{b}
  |\Omega_{m}|^2 < \frac{4\omega_{m}^{2} + (\kappa_3 + \eta^2\kappa)^2}{16\eta^2}.
\end{equation}

For Eq. (\ref{b}), When $\eta = \eta_{min} \equiv \sqrt[4]{4\omega_{m}^{2} + \kappa_{3}^{2}}/\sqrt{\kappa}$, the right of it has minimum $S_{min} = \frac{\kappa}{4}\sqrt{\omega_{m}^{2} + \frac{\kappa_{3}^{2}}{4}} + \frac{\kappa\kappa_{3}}{8}$. Comparing $S_{min}$ with the right of Eq. (\ref{a}), one finds that $S_{min}$ is larger than the right of Eq. (\ref{a}). It indicates that, in comparison to the single cavity, the coupled cavities system tolerates lager optomechanical coupling to keep the system in stable regime.

\section{Conclusion}\label{5}

In conclusion, we have theoretically investigated the ground-state cooling of an optically levitated nanosphere in the high unresolved regime, by introducing a coupled cavity. The auxiliary cavity is coupled with the optomechanical cavity, but does not interact with the levitated nanosphere. This specific configuration of energy transition causes the quantum interference, which modifies the optomechanical response of the nanosphere and gives rise to asymmetry between heating and cooling processes. By tuning the detuning between optomechanical cavity and cooling field, one can take advantage of this interference to enhance the cooling process and restrain the heating process. So that, a larger net cooling rate is obtained in a wide range of parameters and the cooling limit is lowered dramatically. It is found that, ground-state cooling can still be achieved for large optomechanical cavity decay rate $\kappa$ even if the effective optomechanical coupling $\Omega_m$ is weaker than the frequency of the nanosphere $\omega_m$. The cooling limit in our research is sensitive to the the radius of the nanosphere as well as the damping rate of the auxiliary cavity. The increase of nanosphere radius will made the classic cooling limit increase too sudden to remain the nanosphere in ground state regime. The larger the decay rate of auxiliary cavity is, the smaller the optomechanical cavity dissipation that the ground state cooling can tolerate is. The effective interaction between the auxiliary cavity and the levitated nanosphere brings the system from the highly unresolved-sideband regime to an effective resolved-sideband regime. This significantly relaxes the restricted condition that the system must be in the resolved-sideband regime for the nanosphere cooling. Furthermore, the interaction refines the dynamical stability compared to the case without auxiliary cavity. This work may provide the possibility for the corresponding research and application of the levitated nanosphere system beyond the restriction for the current experiment.

\end{document}